\documentstyle[12pt]{article}
\newcommand{\eq}[1]{(\ref{#1})}
\newcommand{\diff}{\partial}
\newcommand{\beq}{\begin{equation}}
\newcommand{\eeq}{\end{equation}}
\newcommand{\beqn}{\begin{eqnarray}}
\newcommand{\eeqn}{\end{eqnarray}}
\newcommand{\cD}{{\cal D}}

\def\cA{{\cal A}}
\def\cC{{\cal C}}

\def\cB{{\cal B}}
\def\cL{{\cal L}}

\def\NP{ Nucl.~Phys.}

\def\PL{ Phys.~Lett.}

\begin{document}

\hfill{ITEP-TH-22/00}

%\hfill{hep-th/mmyynnn}

\vspace{30mm}

\centerline{\bf \Large D-brane annihilation, renormalization-group flow and
non-linear}
\centerline{\bf \Large $\sigma$-model for the ADHM construction}

\vspace{5mm}

\centerline{E.T.Akhmedov\footnote{akhmedov@physics.ubc.ca}}

\vspace{5mm}

\centerline{Institute of Theoretical and Experimental Physics}

\centerline{117259, Moscow, ul. B.Cheremushkinskaya, 25}

\vspace{5mm}

\centerline{and}

\vspace{5mm}

\centerline{University of British Columbia}

\centerline{6224, Agricultural Rd., Vancouver, BC, Canada, V6T 1Z1}

\vspace{10mm}

\begin{abstract}
In this note $D9$- and anti-$D9$-brane annihilation in type I string
theory is probed by a $D1$-brane. We consider the covariant Green-Schwarz
or twistor formulation of the probe theory.
We expect the theory to be $\kappa$-invariant after the annihilation is completed. 
Conditions of the $\kappa$-invariance of the theory impose constraints on the
background tachyon field.  Solutions to the
constraints define tachyon values which correspond to type I $D5$-branes as
remnants of the annihilation.  As a byproduct we get a theory which lies in the
same universality class as the non-linear $\sigma$-model for the
Atiyah-Drinfeld-Hitchin-Manin construction.  \end{abstract}

\vspace{10mm}

\section{Introduction}

 Recently a powerful apparatus to work with BPS excitations in superstring theory
 has been developed (see \cite{Polchinski,reviews} for some reviews).  Despite this progress, we still
have a poor understanding of the dynamics when SUSY is violated. In
fact, even in the simplest situation of $D$-anti-$D$-brane ($D-\bar{D}$)
systems we know only the details of their topological content rather than
the dynamics of the annihilation process \cite{Sen98,Wit99,Hor}.

   First, it is argued that the annihilation process of a
$D-\bar{D}$-system is related to the tachyon rolling down to the bottom of its
energy functional \cite{Sen98}. The tachyon field here describes the lowest energy excitations
of strings stretched between the $D$- and $\bar{D}$-branes \cite{Sen98}.
This excitation has an imaginary mass which signals an instability in the $D-\bar{D}$-system
and leads to the annihilation. Second, knowing which charges are excited, one could trace what kinds of branes
should be remnants of the annihilation \cite{Sen98,Wit99,Hor}.  It is not clear, though,
how to determine their position in their moduli space after the
annihilation has happened.  Moreover, it is not known
how the tachyon rolls down to the bottom of its energy functional during the
annihilation process.  The main obstacle is the lack of 
knowledge of the tachyon potential \cite{tacpot}.

   Rather than looking for the tachyon potential in this note we would like
to implement another approach.  We know that in many occasions $D$-branes
supply a good microscopic description of various low energy physics
phenomena. In particular, the closest example to our approach is  presented in
ref. \cite{Dou95}. In this paper the $D$-brane description of instantons in SUSY Yang-Mills
(SYM) theories is given.  Concretely, the gauge connection
corresponding to the YM instanton is recovered from a microscopic theory
which describes a $D1$-brane in the $D5$-brane background in type I
string theory.  What is most important for us is that the $D1$-brane theory in
question is absolutely restricted by its (4,0) SUSY
invariance \cite{Wit94}.

   Inspired by these considerations, we would like to probe a 
$D9$-$\bar{D9}$-brane annihilation by a $D1$-brane in type I string
theory.  Rather than dealing with the light-cone action for the $D1$-brane \cite{Polchinski} we
consider its covariant Green-Schwarz (GS) formulation
\cite{AgPoSc96}.  For the theory to be non-anomalous it is necessary to consider the
number of $D9$-branes to be the number of $\bar{D9}$-branes plus 32
\cite{Lif,GiPo96,Sug99}. 

  The annihilation process is viewed on the $D1$-brane as a
renormalization-group (RG) flow \cite{Kachru}.  
In fact, after the annihilation some of the strings stretched
between the $D1$-brane and the $D9-\bar{D9}$-system should become
massive and decouple from the IR limit of the probe theory. In the limit the theory
describes the $D1$-brane in the background of only 32 $D9$-branes 
with some gauge bundle on the latter.

  Here we study the theory on the probe brane which
is already a low energy intermediate step in the RG evolution. It is an
approximation to an as yet unknown microscopic theory which contains both $D$-
and $\bar{D}$-branes. Thus, we do not expect to recover from our probe theory
explicitly the way the tachyon rolls down to the bottom of its energy functional. However, we
could hope to extract from it some information about tachyon classical
solutions which respect SUSY, i.e. tachyon values after the annihilation.
In particular, one would hope that there is some symmetry which restricts
possible background values of the tachyon after the annihilation \cite{Anton}. In fact, before 
the annihilation there
is some non-SUSY theory which describes the $D1$-brane in the presence of both $D9$-
and $\bar{D9}$-branes. Via RG flow the theory evolves to a superconformal limit with
a proper background value of the tachyon. We believe that there
should be some hidden ({\it non-linearly} realized) SUSY of the theory,
which forces it to flow in such a rigid way \cite{Anton}.

  It is at this point that $\kappa$-invariance comes into the game. In fact, as
is well established \cite{Sor98,Sezgin}, this symmetry is related to a {\it linearly} realized
SUSY on the world-sheet:  one could formulate the $D1$-brane theory with
an explicit SUSY both on the world-sheet and in the target space.  Then
$\kappa$-symmetry appears from the world-sheet SUSY after the integration
over auxiliary fields \cite{Sor98}. Hence, the presence of the $\kappa$-invariance
is a sign that there is a linearly realized SUSY on the $D1$-brane world-sheet. 
In our case, we expect the invariance to be
present only after the annihilation is completed.

   It is for this reason that we are looking for backgrounds for the
$D1$-brane, which respect $\kappa$-invariance.  Conditions that
invariance imposes on the theory constrain the possible values of the tachyon field.
We find some equations which establish that this field is covariantly constant on 
light-like surfaces in the target space. The latter should be
supplemented by integrability conditions so that if there is a gauge field
background turned on the $D9$- and $\bar{D9}$-branes, the tachyon could be a
non-trivial field rather than just a constant.

   Let us explain why we have equations
for the tachyon field which are linear rather than quadratic 
in differentials. In fact, $\kappa$-invariance of a superstring theory
in a background of SYM fields puts the latter on mass-shell, i.e. one gets
second order differential (classical) equations (of motion) for
the fields \cite{Wit86,Sor98}. Hence, the appearance of the first order differential equation for
the tachyon field might seem suspicious. As we already mentioned, however, the
$\kappa$-invariance is related to SUSY of the world-sheet theory.  Also
the mere presence of the tachyon field explicitly violates
SUSY in the theory. Thus, we expect SUSY to be {\it linearly realized} only for some
specific tachyon values.  That is the reason we get BPS like linear differential
equations for the tachyon field.

   Our main interest is in a soliton which is of co-dimension four within the $D9$-branes.
Only in such a situation is there a SUSY vacuum in the probe theory \cite{Polchinski}.
In this case, the tachyon and background gauge field are functions of
only four coordinates rather than ten. Then the integrability condition in question is just
the self-duality equation for the gauge field. At large distances
its instanton solution is represented as a pure gauge. The gauge matrix
in the latter is equal to a non-trivial map (of a degree equal to the
instanton charge) from $S^3$ at infinity to the group of
Chan-Paton (CP) indexes.  Specifically, we take the target of the map in
question to be the diagonal $USp(4k)$ subgroup of the $USp(4k)\times
USp(4k)$ group.

   This choice could be clarified as follows. First, it should be stressed
that we are considering a minimal construction of $D5$-branes from a
$D9$-$\bar{D9}$-system. This construction is related to that due to Atiyah, Drinfeld, Hitchin and Manin (ADHM)
for instantons in YM theories \cite{ADHM}. In principle, one could study other situations which lead
to non-minimal generalizations of the ADHM construction \cite{Anton}. For "minimality"
we consider a background gauge field on the $D9-\bar{D9}$-brane system 
respecting only the $USp(4k)\times
USp(4k)\times SO(32)$ subgroup of the largest possible group with the same
number of CP indexes. Second, we consider the symplectic groups as
factors because we are looking for a minimal construction which leads to $k$
type I $D5$-branes. As is established in ref. \cite{GiPo96,Wit96} each of the branes should
have two CP indexes, taking values in $USp(2) \cong SU(2)$. Hence, $k$ type I
$D5$-branes correspond to $USp(2k)$ group \cite{GiPo96,Wit96}.  Third, we consider $4k$ rather than
just $2k$ indexes, because we have to embed two of the CP indexes of both $USp(4k)$'s 
into the tangent bundle of the target space (see
\cite{GrScWi} for such a construction).

  Now if the tachyon is covariantly constant in the background in
question it contains the aforementioned map.  Hence, the tachyon field
is a rectangular matrix ($[4k]\times [32+4k]$) whose quadratic part ($[4k]\times [4k]$)
is the map in question.  This is exactly the tachyon value we
expect to get for $k$ type I $D5$-branes to appear as remnants of the
$D9$-brane annihilation \cite{Wit99,Hor}.

   In conclusion, we have a $D9-\bar{D9}$-system with some excited Ramond-Ramond (RR) field
corresponding to $k$ $D5$-branes before the annihilation.  The RR field is encoded in terms of some gauge bundle on the
$D9$-$\bar{D9}$-system \cite{Wit99,Moore}. After
the annihilation, the tachyon acquires a value which is covariantly constant on 
light-like surfaces in the gauge field background.  In our case, such a
tachyon value is proportional to the ADHM matrix \cite{ADHM}, which corresponds to
the ADHM construction of instantons for the $SO(32)$ group. Similarly as in
ref. \cite{Wit94,Dou95}, this matrix defines a mass term for the fermionic
fields on the probe $D1$-brane.

   In this way, we obtain the probe theory which, on the level of massless modes,
coincides with the non-linear $\sigma$-model for the ADHM
construction \cite{Wit94,Anton}. In other words, it flows in the IR limit to the same superconformal
theory which describes the $D5$-brane background for the type I $D1$-brane
theory as an instanton field of the $SO(32)$ group \cite{Strominger}. This
time it is the latter field which encodes the information about corresponding RR charge. The former
gauge field from the diagonal subgroup of $USp(4k)\times USp(4k)$, being a pure
gauge at low energies (due to the non-zero tachyon vacuum expectation value
(VEV)), decouples after the IR limit is taken.

   Thus, without knowing the tachyon potential we could fix tachyon
values after the annihilation.  This is our main result.

\section{Twistor formulation of the probe theory and $\kappa$-invariance}

 We consider a phase of type I string theory containing $32+4k$
$D9$-branes and $4k$ $\bar{D9}$-branes. The D-brane world-volumes fill the entire
ten-dimensional space-time.  We probe the annihilation of the
$D9-\bar{D9}$-system by a $D1$-brane.

  In the GS or twistor formalism the probe theory contains the following 
fields at low energies. First, there are low-energy modes of strings attached by both their ends
to the $D1$-brane.   The modes are ten bosons $x_{M}, \quad (M = 0,..,9)$ and
ten-dimensional Majorana-Weyl fermions $\psi_\cA, \quad (\cA=1,...,16)$
\cite{AgPoSc96,PoWi95}.  Second, there are low-energy modes of strings
stretched between the $D1$- and $D9$-branes. These modes are two-dimensional
Majorana-Weyl fermions $\lambda$ \cite{PoWi95}. Third, there are  also modes
of strings stretched between the $D1$- and $\bar{D9}$-branes.
Correspondingly these modes are two-dimensional Majorana-Weyl fermions $\chi$ of
opposite to $\lambda$ chirality \cite{Sen98,Wit99}. If it were not for
the presence of $\chi$, the $D1$-brane would have the same quantum
numbers as the Heterotic $SO(32)$ string \cite{PoWi95}.

  We are going to work with the twistor formulation of the theory \cite{Sor98,Sezgin}:

\beqn
S = \int d^2\sigma \left\{P_M^{--}\left[e_a^{++}\left(\diff^a x^M
\phantom{\frac12^{\frac12}} -
\phantom{\frac12^{\frac12}} \diff^a
\psi^{\cA}\Gamma^M_{\cA\cB} \psi^{\cB}\right) - \varphi_- \Gamma^M \varphi_-
\right] + \right. \nonumber \\ + {\rm Wess-Zumino \quad term} + \nonumber \\ + \lambda^p
\left[e^a_{--}\left(\delta^{pq}\diff_{a} \phantom{\frac12^{\frac12}}
- \phantom{\frac12^{\frac12}} \diff_{a}x^M A_M^{pq}
\left(x\right)\right) + \frac{1}{4} F^{pq}_{ML}\left(x
\right)\Gamma^{ML}_{\cA\cB} \psi^{\cA} \psi^{\cB}\right]\lambda^q + \nonumber
\\ \left. +
\chi^{\bar{p}} \left[e^a_{++}\left(\delta^{\bar{p}\bar{q}} \diff_a \phantom{\frac12^{\frac12}} -
\phantom{\frac12^{\frac12}}
\diff_{a}x^M B_M^{\bar{p}\bar{q}} \left(x\right)\right) + \frac{1}{4}
H^{\bar{p}\bar{q}}_{ML}\left(x \right)\Gamma^{ML}_{\cA\cB} \psi^{\cA}
\psi^{\cB} \right]\chi^{\bar{q}} + \lambda^p\chi^{\bar{p}} T^{p\bar{p}}
\left(x\right) \right\}.  \label{start}
\eeqn
Here $P_M^a$ and $\varphi_-$ are auxiliary fields. Their exact definition
is not relevant for our further discussion and can be found in
\cite{Sor98}.  These auxiliary fields should obey a Cartan-Penrose
\cite{Sor98} condition:  $P^M_{--} = e_{--}^{-4}\varphi_- \Gamma^M
\varphi_-$.  Now it is easy to see how after the integration over $P^M$ one
recovers the standard GS formulation of the Heterotic string if $\chi$ is absent
\cite{Sor98}.

  Also in this formula $e^a, \, a=1,2$ is a zweibein; $p = 1, ...,
32+4k$ and $\bar{p} = 1, ...,4k$; $\Gamma_M$ ($\Gamma^{ML} = \left[\Gamma^M,
\, \Gamma^L\right]$) are ten-dimensional $\gamma$-matrices in the
Majorana-Weyl representation; $T$ is a tachyon field which describes the lowest
energy excitations of strings stretched between the $D9$- and
$\bar{D9}$-branes \cite{Sen98}.  It appears in \eq{start} as an external
field and transforms in the bi-fundamental representation under the
$USp(4k)\times SO(32)$ and $USp(4k)$ groups. We
choose the gauge fields $A_M^{pq}$ and $B_M^{\bar{p}\bar{q}}$ on the $D9$-
and $\bar{D9}$-branes (with the field strengths $F^{pq}_{ML}$ and
$H^{\bar{p}\bar{q}}_{ML}$, correspondingly) respecting only this subgroup of
the largest possible group with this number of CP indexes.  These gauge
fields couple to $\lambda$ and $\chi$ in the same way as a gauge field couples
to Heterotic fermions \cite{PoWi95}.  All other fields on the $D9$- and
$\bar{D9}$-branes are set to zero.

  Hence, the theory we are starting with is a non-SUSY two-dimensional
$\sigma$-model.  It evolves via the RG flow \cite{Kachru} to a superconformal
theory in the IR if a proper background value of $T$ is standing in
\eq{start} \cite{Anton}.  In fact, some of the $D9$- and $\bar{D9}$-branes should
annihilate leaving only the $D1$-brane in type I string
theory, which contains only 32 $D9$-branes and, possibly, some non-trivial bundles on the latter. 
This $D1$-brane has the quantum numbers of the Heterotic string
\cite{PoWi95} and its theory is superconformal. Thus, one can be sure that if 
the theory in question eventually evolved to a superconformal
limit for some value of $T$, this value really corresponds to a minimum of the tachyon energy
functional.

  As we explained in the introduction, the $\kappa$-invariance could help find
tachyon classical solutions respecting SUSY.  Let us, hence,
impose conditions on the invariance of the action \eq{start} under
$\kappa$-transformations. In the conformal gauge the transformations look as follows
\cite{Dha86,Sezgin,Sor98}:

\beqn
\delta P_M^a = 0, \quad \delta\varphi_- = 0 \nonumber \\
\delta\psi^{\cA} = 2i P^M_{--} \Gamma_M^{\cA\cB} \kappa_{\cB++} \nonumber
\\ \delta x^M = -i \delta\psi^{\cA}\Gamma^M_{\cA\cB}\psi^{\cB} \nonumber \\
\delta \left(A_M\diff_{--} x^M\right) =
\diff_{--}\Lambda_{\kappa} + \left[\Lambda_{\kappa},
A_M\diff_{--} x^M\right], \quad \delta\lambda^{p} =
\left(\Lambda_{\kappa}\lambda\right)^p  \label{trans}
\eeqn
where $\Lambda_{\kappa} = \delta x^M A_M$ if only the background {\it gauge} field is
non-zero. Also, by analogy with the transformations of $\lambda$ we could
choose a natural transformation law \cite{Dha86} for $\chi$ to be:

\beq
\delta \left(B_M\diff_{++} x^M\right) =
\diff_{++}\Lambda'_{\kappa} + \left[\Lambda'_{\kappa},
B_M\diff_{++} x^M\right], \quad \delta\chi^{\bar{p}} =
\left(\Lambda'_{\kappa}\chi\right)^{\bar{p}},
\eeq
and $\Lambda'_{\kappa} = \delta x^M B_M$. At the same time the tachyon field
transforms under the $\kappa$-symmetry simply as follows:

\beq
\delta T \left(x\right) = \diff_M T \left(x\right) \cdot \delta x^M
\eeq

  It is necessary to supplement these transformations by Virasoro and SYM
constraints \cite{Wit86,Sor98}. In our case, the latter are equivalent to the classical SYM
equations of motion.  Moreover, for \eq{start} to be invariant under
\eq{trans} the tachyon field $T$ should obey the following equation:

\beqn
\delta x^{\cA\cB}\cdot \hat{\cD}_{\cA\cB} T\left(x\right) = \nonumber \\ = \delta
x^{\cA\cB}\cdot \left\{\hat{\diff}_{\cA\cB} T^{q\bar{p}}\left(x\right) +
\phantom{\frac12^{\frac12}}
T^{q\bar{q}}\left(x\right) \cdot
\hat{B}^{\bar{q}\bar{p}}_{\cA\cB}\left(x\right) \phantom{\frac12^{\frac12}} -
\hat{A}_{\cA\cB}^{qp}\left(x\right) \cdot T^{p\bar{p}}\left(x\right)\right\} =
0 \nonumber \\ {\rm where} \quad \delta x^{\cA\cB} = \hat{P}_{--}^{\cA\cC}
\cdot \kappa_{\cC++} \cdot \psi^{\cB} \quad {\rm and} \quad \hat{P}_a^{\cA\cC}
= P_a^M \Gamma_M^{\cA\cC} \label{statement}
\eeqn
for {\it any} $\kappa$.  Hence, this should be supplemented by
integrability conditions:

\beq
\left[\hat{P} \cdot\kappa_{(1)}\cdot\psi\cdot \hat{\cD}, \phantom{\frac12^{\frac12}}
\hat{P} \cdot\kappa_{(2)}\cdot\psi\cdot\hat{\cD}\right] = \delta x^{(1)}
\cdot \delta x^{(2)} \cdot \left[\cD^{(1)}, \, \cD^{(2)}\right] =
0. \label{integrab}
\eeq
Note that the vector $\hat{P} \cdot \kappa \cdot \psi$ has zero
norm:$\left(\hat{P}\cdot\kappa\cdot\psi\right)^2 \sim P^2_{--} \cdot
\epsilon_{\cA\cB} \psi^{\cA}\psi^{\cB} \cdot \\ \cdot \delta^{\cC\cD}
\kappa_{\cC++}\kappa_{\cD++} = 0$ due to the anti-commutativity of $\kappa$.
Hence, $\hat{P} \cdot \kappa \cdot \psi$ is a constant, (independent of $x_M$)
light-like vector in the ten-dimensional Minkowski space. Moreover, under
variations of the parameter $\kappa$ (with $P$ and $\psi$ kept fixed) it sweeps an
eight-dimensional hyperplane in the ten-dimensional space-time. In fact, as is
well known, the matrix $\hat{P} \cdot \kappa$ has eight rather than sixteen
non-zero eigen-values \cite{GrScWi}. Hence, it defines eight real deformations of $\kappa$.
Thus, there are eight varying components of the ten-vector in question, while
the other two are fixed. This is the eight-dimensional hyperplane. At the same time
under variations of $\hat{P}$ and $\psi$ all eight-dimensional hyperplanes are swept.

\section{Solutions to the constraints and $D5$-branes as remnants
of the annihilation}

  As follows from \eq{integrab} when the
background $A_M$ and $B_M$ are zero, the tachyon field should be a constant up to a
gauge transformation.  Unfortunately we can not derive from our formulae what
kind of constant it should be.  However, as a warm up exercise let us try to
guess it \cite{Anton}.  Consider:

\beq
T_{[4k + 32]\times [4k]} \sim
\left(D_{[4k]\times[4k]} \oplus 0_{[32]\times[4k]}\right), \label{VEV1}
\eeq
where $D$ is a diagonal matrix with all eigen-values of the order of the
string scale. This tachyon value respects both $SO(32)$ and the diagonal
$USp(4k)$ subgroups of the $SO(32)\times USp(4k)\times
USp(4k)$ group.  (See \cite{SenD,Yi} for discussion on this subject.)
Gauge invariant expression for the tachyon VEV should be:

\beq
T^{\bar{p}p} \cdot T_{p}^{\bar{q}}
= \delta^{\bar{p}\bar{q}}. \label{VEV2}
\eeq
Presumably the latter expression is
related to the minimum of the tachyon potential: $\diff_T V(T)|_{T^2 = 1} =
0$, but its origin is not really important to us. 

   What is most important is that the formula \eq{VEV2} passes through the simplest check. In fact, consider
the lowest energy excitations in the NS sector of the strings stretched between the $D1$-brane
and the $D9-\bar{D9}$-system  \cite{Polchinski}. We denote these excitations as $Q^p$ and $\tilde{Q}^{\bar{p}}$, respectively. Their
bare masses are equal to $\frac12$ in string units \cite{Polchinski}.
Also their interactions with the tachyon field are
$T^{p\bar{p}}\cdot T_{\bar{p}}^q \times Q_p \cdot Q_q$ and 
$T^{\bar{p}p} \cdot T_{p}^{\bar{q}} \times \tilde{Q}_{\bar{p}}\cdot\tilde{Q}_{\bar{q}}$. Hence, when the tachyon
acquires the VEV as in \eq{VEV1},\eq{VEV2} we have the proper number of the NS modes
with mass $\frac12$ to describe the $D1$-brane in the background of 32 $D9$-branes only. 
We are going to use eq. \eq{VEV2} later.

   With the tachyon as in the eq. \eq{VEV1}, the $\chi^{\bar{p}}$ and
$\lambda^{\bar{p}}$ from \eq{start} become massive, while $\lambda^n$, $n=1,...,32$ are left massless:
here $\lambda^p = (\lambda^{\bar{p}}, \lambda^n)$.
Because of IR effects in two dimensions, if such fields acquire masses there
is no way for them to become massless.  Hence, in the study of the RG evolution of the
theory we can safely integrate these massive fields out, while leaving
massless ones untouched. After the integration, we get the ordinary type I $D1$-brane theory. 
This theory describes the background of only 32 $D9$-branes and is superconformal \cite{PoWi95}. Thus,
\eq{VEV1} is a proper VEV for the tachyon in this case. We just
guessed the VEV in question but in the situation below we derive it.

 Now let us study the case when there is a co-dimension four
soliton left within the $D9$-branes after the annihilation. This corresponds to a
solution of eq. \eq{integrab} when $A_M$, $B_M$ and $T$ are
functions of four coordinates. Say the latter are $x_6,...,x_9$.
In this case we expect a linear realization of SUSY in the
IR limit\footnote{We suppose, but can not prove, that in all other situations eq. \eq{statement}
and \eq{integrab} have trivial solutions.}.

   In the presence of the soliton in question, the ten-dimensional Lorentz invariance is broken: $SO(9,1)\to
SO(1,1)\times SO(4) \times SO(4)$. Here $SO(1,1)$ corresponds to rotations
along the $D1$-brane (directions\footnote{From now on we fix the light-cone gauge.} $0,1$); one of the $SO(4)$'s is related to
rotations along the soliton, but transverse to the $D1$-brane (directions
$2,...,5$), while another $SO(4)$ corresponds to rotations in the
directions transversal to the soliton ($6,...,9$).

    Below we denote by $\pm$ the left and right chirality under the $SO(1,1)$ group.  At the same time by
$\alpha$ and $\dot{\alpha}$ we denote the indexes of the fundamental
representation of the $SU(2)_L$ and $SU(2)_R$ subgroups of the aforementioned
``transversal'' $SO(4)$ group.  Besides that we embed two among the CP indexes $p$
and $\bar{p}$ into the tangent bundle of the target space.  Thus,
$p=(\alpha,i,n)$ and $\bar{p}=(\dot{\alpha},i)$, where $n=1,...,32$ and
$i=1,...,2k$: hence, both $USp(4k)$'s are broken to $SU(2)\times USp(2k)$.
In other words, the
fermions on the $D1$-brane carry the following indexes $\chi^{\bar{p}} = \chi_-^{\dot{\alpha}i}$
and $\lambda^p = \left(\lambda_+^{\alpha i}, \lambda_+^n\right)$.

   If $A_M$, $B_M$ and $T$ are functions of the four coordinates only, then
\eq{statement} and \eq{integrab} take the form:

\beq
d^{\alpha\dot{\alpha}}_{(1),(2)} \cD_{\alpha\dot{\alpha}}
T\left(x\right) = 0 \quad {\rm and} \quad
d^{\alpha\dot{\alpha}}_{(1)} d^{\beta\dot{\beta}}_{(2)}
F_{\alpha\dot{\alpha}\beta\dot{\beta}}\left(x\right) = 0, \label{fourD}
\eeq
where $F_{\mu\nu} = \left[\cD_{\mu},\cD_{\nu}\right], \quad
F_{\alpha\dot{\alpha}\beta\dot{\beta}} = F_{\mu\nu} \cdot
\tau^{\mu}_{\alpha\dot{\alpha}} \cdot \tau^{\nu}_{\beta\dot{\beta}}$, $\mu =
6,...,9$. Also here $d_{\alpha\dot{\alpha}}^{(1),(2)}$ are constant
(independent of $x_{\mu}$) vectors with zero norm in the complexified
four-dimensional Euclidean space. They correspond to $\hat{P} \cdot \kappa
\cdot \psi$ with two different $\kappa$'s.

   We consider complexification of the Euclidean space (use complex
$x_{\alpha\dot{\alpha}}$ coordinates rather than real $x_{\mu}$) and
complexify the gauge group to show that there are non-trivial solutions to
the eq.  \eq{fourD}.  What is most important, if $\hat{P}$ and $\psi$ are fixed,
the vector $d_{\alpha\dot{\alpha}}$ sweeps a
complex two-plane ($\beta$-plane in the notation of ref. \cite{Ward}) in the complex four-dimensional
space: In the case under
study $\hat{P}\cdot\kappa$ describes two complex deformations. At the same time, under variations
of $\hat{P}$ and $\psi$ the vector $d_{\alpha\dot{\alpha}}$ sweeps all light-like
surfaces in the space.

   Before going further we would like to remind that we are
looking for a minimal construction of the $D5$-branes out of the
$D9$-$\bar{D9}$-system.  As we mentioned in the introduction, to construct $k$ type I
$D5$-branes as a result of a $D9$-brane annihilation we need $4k+32$
$D9$-branes and $4k$ $\bar{D9}$-branes. Besides that we take $A_{\mu}^{nm}$
to be a pure gauge.  Then, the second equation in \eq{fourD} is equivalent
\cite{Ward} to the self-duality condition for the YM connection
$\tilde{A}_{\mu}^{\bar{p}\bar{q}} = A_{\mu}^{\bar{p}\bar{q}} -
B_{\mu}^{\bar{p}\bar{q}}$ from the diagonal subgroup of
$USp(4k)\times USp(4k)$. 

   To find IR limit of the theory \eq{start} we
need to know the large distance behavior of the instanton solution to eq. 
\eq{fourD}.  With the charge $2k$ the solution behaves at infinity as
follows:

\beqn
\tilde{A}^{\bar{p}\bar{q}}_{\mu} \sim \left(\left(\diff_{\mu} \hat{S} \right)
\phantom{\frac12^{\frac12}} \hat{S}^{-1}\right)^{ij}_{\alpha\beta} \quad {\rm
where} \quad S_{\alpha\dot{\alpha}}^{ij} = \delta^{ij} \cdot
\frac{\left(\hat{x}_{\dot{\alpha}\alpha} - \hat{x}^{(i)}_{\dot{\alpha}\alpha}
\right)}{\left|x - x^{(i)}\right|}, \quad |\hat{x}|\to\infty. \label{gaugf1}
\eeqn
Here $\hat{x}_{\alpha\dot{\alpha}} = x_{\mu}\tau^{\mu}_{\alpha\dot{\alpha}}$
and $x_{(i)}$ are positions of the $2k$ instantons.  This is not the most
general behavior at infinity but we use it to clarify our idea.

   The gauge field \eq{gaugf1} defines a map $\hat{S}$ of the order
$2k$ from $S^3 \cong USp(2) \cong SU(2)$ at space infinity to the
diagonal subgroup of $USp(4k)\times USp(4k)$.  In fact:

\beq
\left(\left(\diff_{\mu} \hat{S}\right) \phantom{\frac12^{\frac12}}
\hat{S}^{-1} \right)^{ij}_{\alpha\beta} = 2\sigma_e^{\alpha\beta}\cdot
\eta^e_{\mu\nu} \cdot \delta^{ij} \cdot \frac{\left(x -
x^{(i)}\right)_{\nu}}{\left|x - x^{(i)}\right|^2}, \quad e = 1,2,3 \, ,
\eeq
where $\eta^e_{\mu\nu}$ are t'Hooft symbols and $\sigma_e$ are generators of
the $SU(2)$ group.  This is just a singular instanton of the $USp(4k)$ group
with the charge $2k$.

  Plugging \eq{gaugf1} into \eq{fourD}, we
find that the tachyon field behaves at large distances as:

\beq
T^j_{p\dot{\alpha}} \sim \left\{\frac{\left(\hat{x}_{\dot{\alpha}\alpha} -
\hat{x}^{(i)}_{\dot{\alpha}\alpha} \right)}{\left|x -
x^{(i)}\right|}\delta^{ij} \oplus 0^{in}_{\dot{\alpha}}\right\}, \quad |\hat{x}|\to\infty. \label{Btach}
\eeq
This is the tachyon value found in \cite{Sen98,Wit99}. It describes $k$ singular
$D5$-branes within type I string theory after the annihilation.

   Now let us consider the more general situation, i.e. deform the gauge field
\eq{gaugf1} to:

\beqn
\tilde{A}^{\bar{p}\bar{q}}_{\mu} \sim \left(\left(\diff_{\mu} \hat{S}
\right)\phantom{\frac12^{\frac12}} \hat{S}^{-1}\right)^{ij}_{\alpha\beta},
\quad {\rm where} \quad S_{\alpha\dot{\alpha}}^{ij} =
\Delta^{il}\left(x\right)\cdot \left(\hat{x}_{\dot{\alpha}\alpha} \delta^{lj}
- \hat{X}^{lj}_{\dot{\alpha}\alpha} \right), \quad |\hat{x}|\to\infty  \nonumber \\ {\rm and} \quad
\left(\Delta^{-2}\right)^{ij} = \left\{\left(x \delta^{il} -
X^{il} \right)_{\mu}\left(x \delta^{lj} -
X^{lj} \right)_{\nu} \phantom{\frac12^{\frac12}} -
\phantom{\frac12^{\frac12}}\left[X_{\mu},\,
X_{\nu}\right]^{ij}\right\}\tau^{\mu}\tau^{\nu}.  \label{gaugf}
\eeqn
Here $\hat{S}$ defines a most general (up to gauge transformations) map of
the order $2k$ from $S^3$ at spatial infinity to the diagonal subgroup of
$USp(4k)\times USp(4k)$. In this formula $\hat{X}^{ij}_{\alpha\dot{\alpha}}$
is an arbitrary symplectic matrix from the diagonal subgroup of
$USp(2k)\times USp(2k)$, which obeys the reality condition
$\hat{X}^{\alpha\dot{\alpha}} =
\epsilon^{\alpha\beta}\epsilon^{\dot{\alpha}\dot{\beta}}
\hat{X}^*_{\beta\dot{\beta}}$.

  With this value of the gauge field substituted into eq. \eq{fourD}, we find
that the tachyon behaves at infinity as \cite{Anton}:

\beq
T^j_{p\dot{\alpha}} \sim \Delta^{il} \cdot \left\{
\left(\hat{x}_{\dot{\alpha}\alpha} \cdot \delta^{lj} -
\hat{X}^{lj}_{\dot{\alpha}\alpha}\right) \phantom{\frac12^{\frac12}}\oplus
\phantom{\frac12^{\frac12}}
h^{ln}_{\dot{\alpha}}\right\}, \quad {\rm where} \quad h^{\dot{\alpha}jn} =
\epsilon^{ij}\epsilon^{\dot{\alpha}\dot{\beta}}
\left(h^n_{i\dot{\beta}}\right)^*, \quad |\hat{x}|\to\infty. \label{VEVT}
\eeq
It is a solution to eq. \eq{fourD} up to the gauge transformation by the matrix
$\Delta^{ij}$ from the diagonal subgroup of $USp(2k)\times USp(2k)$. Here
$\epsilon^{ij}$ is $USp(2k)$ invariant tensor served to raise and lower $i$
indexes.

  So far $h^{in}_{\dot{\alpha}}$ in eq. \eq{VEVT} is an arbitrary matrix, i.e. not fixed
by the eq. \eq{fourD}.  However, taking into account eq. \eq{VEV2}, or
$T^j_{p\dot{\alpha}}\cdot T^{i\dot{\beta}p} \sim
\delta^{ij}\cdot\delta_{\dot{\alpha}}^{\dot{\beta}}$ in our case,
the matrices $X$ and $h$ should obey the ADHM condition \cite{ADHM}:

\beqn
\left(\epsilon^{\alpha\beta}\hat{X}_{\alpha\dot{\alpha}}
\hat{X}^{\beta\dot{\beta}}\right)^{ij} +
h^{in}_{\dot{\alpha}}h^{nj\dot{\beta}} = 0, \label{condition2}
\eeqn
with such a gauge choice as in eq. \eq{VEVT}.
Note that when $h = 0$ we recover the situation of the singular $D5$-brane
\eq{Btach}.

   Let us now check whether or not we have found a proper tachyon value which
minimizes its energy functional. Specifically we are going to
check whether or not theory \eq{start} flows to a superconformal limit 
in the IR with such a tachyon value.

   Substituting the value \eq{VEVT} for the
tachyon field into the action \eq{start} we get:

\beqn
\cL = \cL_{kin}(x,\psi,\chi,\lambda) + \chi^{\alpha j}_-\cdot
\Delta^{jl}
\cdot \left(\left(\hat{x}_{\dot{\alpha}\alpha}\cdot \delta^{li} -
\hat{X}_{\dot{\alpha}\alpha}^{li}\right)\lambda^{\dot{\alpha} i}_+
\phantom{\frac12^{\frac12}} + \phantom{\frac12^{\frac12}}
h^{ln}_{\alpha} \lambda^n_+\right). \label{prom}
\eeqn
Here we showed spinor indexes to present a close similarity of our theory to
that considered in ref. \cite{Dou95}. To understand how the theory \eq{prom} evolves under the RG
flow, one must find massless fields among $\lambda$ and $\chi$.  For
this purpose it is necessary to look for a complete set of solutions to the
equation \cite{Wit94,Dou95}:

\beq
T^j_{p\alpha}(x) v^{pn}(x) = 0, \label{solTach}
\eeq
for a general $X^{ij}$ and $h^{in}_{\dot{\alpha}}$ obeying \eq{condition2}.
Once we have found all $32$ solutions to these equations, it is possible to
decompose $\lambda_+$ in their basis:

\beq
\lambda^p_+ = \sum^{32}_{n = 1} v^{pn} \lambda^n_+.
\eeq
Substituting this expression into \eq{prom} and integrating out massive
modes, we get:

\beq
\cL = \cL(x,\psi)_{kin} + \lambda^n_+\left(\diff_{--}\delta^{nm}
\phantom{2^{2^{\frac12}}} + \phantom{2^{2^{\frac12}}}
\diff_{--} \hat{x}^{\dot{\alpha}\alpha}
\cdot \hat{A}_{\dot{\alpha}\alpha}^{nm}(x)\right)\lambda^m_+, \label{lag}
\eeq
where $\hat{A}^{nm}_{\dot{\alpha}\alpha}(x) = \left(v^{n}_p\right)^{-1}
\frac{\diff}{\diff x^{\dot{\alpha}\alpha}} v^m_p$. Note also that the
gauge field \eq{gaugf}, being a pure gauge at low energies (due
to the non-zero tachyon VEV \eq{VEV2}),
does not enter the IR Lagrangian \eq{lag}. The
theory \eq{lag} is superconformal \cite{Dou95,Wit94}.

   Taking into account \eq{VEVT},\eq{condition2} and \eq{solTach} we see that
$\hat{A}_{\alpha\dot{\alpha}}^{nm}$ is the self-dual vector-potential
corresponding to the ADHM ``matrices'' $X$ and $h$. This vector-potential
describes $k$ type I $D5$-branes in some non-singular (but otherwise
generic) point of their moduli space \cite{Dou95}.  So, choosing the value of
$T$ as in \eq{VEVT} and \eq{condition2}, we arrive via RG at the
superconformal theory \eq{lag}. It describes the type I $D1$-brane in the
background of $k$ $D5$-branes.

  Now let us discuss a difference between our theory and that considered in
\cite{Dou95}. One immediately sees that the massive spectra of the two
theories are different \cite{Anton}. First, because of the factor $\Delta^{ij}$, masses of
the massive fermions in \eq{prom} are different from those of fermions
in the theory from ref. \cite{Dou95}.
Second, scalars which are present in \cite{Dou95} and correspond (along with
$\chi_-^{\alpha j}$ and $\lambda_+^{\dot{\alpha}j}$) to the strings stretched between
$D1$- and $D5$-branes, are absent in eq. \eq{prom}.  These scalars are
massive at a generic point of the instanton moduli space \cite{Dou95}.

   It is worth mentioning at this point that one should not expect the theory
\eq{prom} to properly reproduce all massive modes.  In fact, this theory is
a low-energy one, because it depends on classical values of macroscopic
fields such as $T$ and $A$ \cite{Anton}. Hence, the theory does not contain 
microscopic degrees of freedom.

\section{Conclusions and Acknowledgments}

 Thus, the theories from eq. \eq{prom} and from ref.
\cite{Wit94,Dou95}, while being different at high energies, flow to the same
superconformal field theory in the IR.  We believe in the existence of a
microscopic theory underlying both of the theories in question \cite{Anton}.  
After the integration of one type of its massive modes, the microscopic theory should lead to the theory considered in
\cite{Dou95} as an intermediate step of the RG flow.  However, the
integration of another type of its massive modes leads the microscopic theory to the Lagrangian \eq{prom} as an
intermediate step of the RG flow.

  I would like to acknowledge discussions with D.Sorokin, D.Polyakov,
A.Sen, N.Berkovits, E.Sezgin, A.Gorsky,
K.Zarembo, A.Rosly and especially with A.Gerasimov. Also I would like to thank
P.Horava for encouraging me to deal with the subject in question. This work was done
under the support of NSERC NATO fellowship grant and under the partial
support of grants INTAS-97-01-03 and RFBR 98-02-16575.


\begin{thebibliography}{50}
\bibitem{Anton} I would like to thank A.Gerasimov for discussions on these
subjects.
\end{thebibliography}
\end{document}